\documentstyle[prl,aps,amsmath,%
               amsfonts,amsopn,graphicx,%
               pstricks]{revtex}

\newpsobject{showgrid}{psgrid}{subgriddiv=1,griddots=10,gridlabels=7pt}

\DeclareMathOperator{\re}  {Re}
\DeclareMathOperator{\im}  {Im}

\DeclareMathOperator{\I}   {i}
\DeclareMathOperator{\mdot}{\mspace{-2.0mu} \cdot \mspace{-2.0mu}}

\newcommand{\ash}{\alpha_{\rm s}}
\newcommand{\qs}{Q_{\mathrm{s}}}

\newcommand{\E}{{\mathrm{e}}}

\newcommand{\abs}  [1]{\lvert#1\rvert}

\newcommand{\sci}  [1]{\times 10^{#1}}

\newcommand{\nabn}{\nabla^2}

\renewcommand{\vec}[1]{{\mathbf{#1}}}

\newlength{\mlen}

\newcommand{\eno}  [1]{\mbox{\!(\ref{#1})}}
\newcommand{\eqn}  [1]{\mbox{Eq.\! (\ref{#1})}}
\newcommand{\eqns} [2]{\mbox{Eqs.\! (\ref{#1})~and~(\ref{#2})}}

\newcommand{\fig}  [1]{\mbox{Fig.~\ref{#1}}}
\newcommand{\figs} [2]{\mbox{Figs.~\ref{#1}~and~\ref{#2}}}

\newcommand{\figa} [2]{\mbox{Fig.~\ref{#1}#2}}
\newcommand{\figab}[4]{\mbox{Fig.~\ref{#1}#3}~and~\ref{#2}#4}

\makeatletter
\def\@eqnacr{{\ifnum0=`}\fi\@ifstar{\@yeqnacr}{\@yeqnacr}}
 
\def\@yeqnacr{\@ifnextchar [{\@xeqnacr}{\@xeqnacr[\z@]}}
 
\def\@xeqnacr[#1]{\ifnum0=`{\fi}\cr \noalign{\vskip\jot\vskip #1\relax}}
 
\def\eqalign{\null\,\vcenter\bgroup\openup1\jot \m@th \let\\=\@eqnacr
\ialign\bgroup\strut
\hfil$\displaystyle{##}$&$\displaystyle{{}##}$\hfil\crcr}
\def\endeqalign{\crcr\egroup\egroup\,}
\makeatother

\begin{document}
\draft

\title{Finite Wavelength Instabilities in a\\
  Slow Mode Coupled Complex Ginzburg-Landau Equation}

\author{M.\ Ipsen}
\address{UNI$\bullet$C, 
  Danish Computing Center for Research and Education,
  The Technical University of Denmark,
  Building~304,
  DK-2800 \mbox{Lyngby},
  Denmark}
\author{P.\ G.\ S{\o}rensen}
\address{Department of Chemistry,
  University of Copenhagen,
  H.C.{\O}rsted Institute,
  Universi\-tets\-parken~5,
  DK-2100 Copenhagen,
  Denmark}

\date{\today}
\maketitle
\begin{abstract}
  In this letter, we discuss the effect of slow real modes in
  reaction-diffusion systems close to a supercritical Hopf
  bifurcation.  The spatio-temporal effects of the slow mode
  cannot be captured by traditional descriptions in terms of a
  single complex Ginzburg-Landau equation (CGLE).  We show
  that the slow mode coupling to the CGLE, introduces a novel
  set of finite-wavelength instabilities not present in the
  CGLE.  For spiral waves, these instabilities highly effect
  the location of regions for convective and absolute
  instability.  These new instability boundaries are
  consistent with transitions to spatio-temporal chaos found
  by simulation of the corresponding coupled amplitude
  equations.
\end{abstract}

\pacs{47.20.Ky, 52.35.Mw}

\label{sec:Intro}
Amplitude equations have been used in a variety of scientific
contexts to describe spatio-temporal modulations of reference
states close to the onset of criticality.  In dynamical
systems close to a supercritical Hopf
bifurcation~\cite{Marsden76}, the spatio-temporal modulation
of the homogeneous stationary state can be described by the
complex Ginzburg-Landau equation (CGLE).  This applies to
chemical and biochemical reaction-diffusion systems, among
which the Belousov-Zhabotinsky (BZ) reaction is the most well
known~\cite{FieldBurger}.

For chemical systems\cite{Kuramoto,Ka95}, one of the most
striking features of the CGLE is its ability to exhibit spiral
wave solutions (point defects) similar to experimental
observations in a large number of chemical reaction-diffusion
systems.  Recently, it was shown~\cite{Trans98} that the CGLE
fails to model even qualitatively the dynamics of a realistic
4-species Oregonator model~\cite{Wang95} of the BZ reaction.
This discrepancy is caused by the presence of a slow real mode
in the homogeneous part of the Oregonator model. By
considering a slow-field coupling to the CGLE, we show how the
inclusion of an amplitude equation for the slow mode gives
rise to a finite-wavelength instability for plane waves which
is not present in the CGLE.  For spiral waves, we have
calculated new boundaries for convective and absolute
instability.

\label{sec:Instabilites}
Here we consider reaction-diffusion systems whose
spatio-temporal dynamics is governed by
\begin{equation}
  \partial \vec{c}/\partial t = 
  \vec{F}(\vec{c};\mu) + \vec{D} \mdot \nabn \vec{c},
  \label{eq:ReacDiff}
\end{equation}
where $\vec{c} = \vec{c}(\vec{x},t)$ depends on the spatial
position vector $\vec{x}$ and time $t$, and $\vec{D}$ is a
diffusion matrix.  Close to the onset of a supercritical Hopf
bifurcation of a homogeneous solution of \eqn{eq:ReacDiff},
the spatio-temporal modulation of this state can be described
by the complex Ginzburg-Landau equation (CGLE).  In
dimensionless form, the CGLE can be written compactly as
\begin{equation}
  \dot{w} = w - (1 + \I\alpha) w\abs{w}^2 + (1 + \I\beta) \nabn w.
  \label{eq:DimCGLE}
\end{equation}
As shown in~\cite{Kuramoto}, the two real coefficients,
$\alpha$ and $\beta$, and the transformation from $w$ to
chemical concentration $\vec{c}$ can be derived rigorously
from the original reaction-diffusion system~\eno{eq:ReacDiff}.
The CGLE admits plane wave solutions of the form $w(t,\vec{x})
= A \exp[\I(\vec{Q x} - \omega t)]$, with amplitude $A =
\sqrt{1-Q^2}$ and frequency $\omega$ determined by the
dispersion relation $\omega = \beta Q^2 + (1 - Q^2) \alpha$
(where $Q = \abs{\vec{Q}}$).  The stability of a given plane
wave is determined by the growth rate $\lambda(k)$ of
perturbations with $\vec{k} \parallel \vec{Q}$
\begin{equation}
  \lambda(k) = -(k^2 + 2\I \beta k Q + A^2)
  \pm \sqrt{(1+\alpha)^2 - k^2 + 2\I \beta k Q + A^2)^2}.
  \label{eq:EigCGLE}
\end{equation}
In particular, at the Eckhaus border defined by
\begin{equation}
  D_\parallel = 1 + \alpha\beta - 2(1+\alpha^2)Q^2/(1-Q^2) = 0,
  \label{eq:Eckhaus}
\end{equation}
a plane wave with given $Q$, will be unstable to
long-wavelength perturbations if $D_\parallel(Q) < 0$;
finally, all plane waves are rendered unstable at the
Benjamin-Feir-Newell (BFN) instability~\cite{New74} where $1 +
\alpha\beta < 0$.

For simple oscillatory chemical systems, the CGLE shows an
almost quantitative agreement with the spatio-temporal
dynamics of the actual chemical system \cite{Ipsen97}.
However, for more complicated models of the BZ reaction, we
have previously described \cite{Trans98} how the CGLE fails
even qualitatively to model characteristic time and length
scales of the models.  This disagreement is caused by the
presence of a slow (near-critical) real mode.  To incorporate
the dynamics of the slow real mode into a description valid
close to criticality, one may derive an amplitude equation
similar to the normal form associated with a fold-Hopf
bifurcation for homogeneous systems~\cite{GuckHolmes}.  In
dimensionless representation, the amplitude equations becomes
\begin{subequations}
  \begin{align}
    \label{eq:DsheA}%
    \dot{w} &= w + (1 + \I\gamma)wz - (1 + \I\ash) w\abs{w}^2 + 
    (1 + \I\beta) \nabn w,\\ 
    \label{eq:DsheB}%
    \epsilon
    \dot{z} &= \lambda_0 z + \kappa \abs{w}^2 + \epsilon\delta \nabn z,
  \end{align}
  \label{eq:DSHE}%
\end{subequations}
where $w$ and $z$ describe the complex and real amplitudes of
the oscillatory and slow real mode respectively.  The
parameter $\lambda_0$ is the reciprocal timescale of the slow
real mode and $\epsilon$ describes the distance to the Hopf
bifurcation point.  The resonant nonlinear coefficients
$\ash$, $\gamma$, and $\kappa$ can be derived by application
of classical normal form theory~\cite{GuckHolmes}.  We shall
refer to the system~\eno{eq:DSHE} as the distributed slow-Hopf
equation (DSHE).  The DSHE may be considered as a ``normal
form'' or prototype model for oscillatory reaction-diffusion
systems with a slow real mode.  It describes, for example, a
realistic 4-species model for the BZ-reaction very
well~\cite{Trans98}.  In the following, we shall use
$\lambda_0 = -3.07\sci{-4}$, $\gamma = -1.56$, $\kappa =
-3.10\sci{-4}$, and $\delta = 0.67$ as calculated for this
model, whereas $\ash$, $\beta$, and $\epsilon$ are regarded as
free parameters.  The value of $\epsilon$ in a realistic
experiment is of the order $10^{-3}$.

Observe that the DSHE~\eno{eq:DSHE} cannot be rescaled to be
fully independent of the distance $\epsilon$ from the Hopf
bifurcation point: Except for a rescaling of the amplitude of
$w$, the DSHE converges to the CGLE when $\epsilon \rightarrow
0$ or $\lambda_0 \rightarrow -\infty$.  In either of these
limits, the coefficient $\ash$ is related to the nonlinear
coefficient $\alpha$ in the CGLE by
\begin{equation}
  \ash = 
  (\alpha + \gamma \kappa/\lambda_0)/
  (1 + \kappa/\lambda_0).
  \label{eq:AlphaDSHE}
\end{equation}
In the adiabatic approximation where either the operating
point is sufficiently close to the Hopf point ($\epsilon$
small) or when the real mode $\lambda_0$ becomes large and
negative, we expect the dynamics of the DSHE to be fully
described within the framework of the CGLE.  However, the
distance from the bifurcation point where the adiabatic
approximation holds may very well be extremely small and very
likely experimentally unrealizable.

The DSHE admits a family of plane wave solutions of the form
\begin{equation}
  w(t,\vec{x}) = A \exp [\I(\vec{Q x} - \omega t)], \qquad
  z(t,\vec{x}) = Z,
  \label{eq:PlaneWave}%
\end{equation}
where $A = \sqrt{(1-Q^2)/(1 + \kappa/\lambda_0)}$, $Z = -A^2
\kappa/\lambda_0$, and the frequency $\omega$ given by the
dispersion relation $\omega = \beta Q^2 + \alpha A^2$ with
$\alpha$ given by \eqn{eq:AlphaDSHE}.

\begin{figure}[htbp]
  \begin{center}
  \begin{pspicture}(0,0.2)(6.7,4.2)
    \rput[br](6.7,0.9){%
      \includegraphics[width=9.5cm]{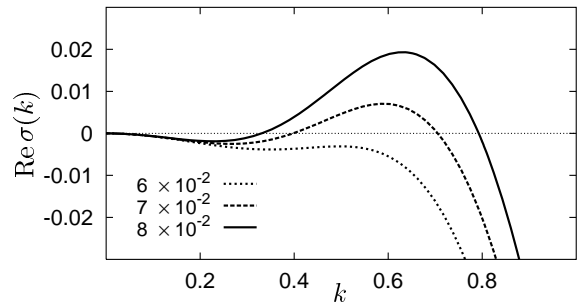}%
      }
  \end{pspicture}
  \caption{%
    Behavior of the real part of the most unstable eigenvalue
    $\sigma(k)$ for a plane wave solution of the DSHE for
    three different parameter values of the parameter
    $\epsilon$ close to criticality ($\alpha = 1.85$ and
    $\beta = -1$).  The slow field coupling in \eqn{eq:DSHE}
    causes a finite-wavelength instability to occur before the
    onset of long-wavelength instabilities.
    }%
  \label{fig:Eigval}
  \end{center}
\end{figure}

\begin{figure}[htbp]
  \begin{center}
    \begin{pspicture}(0,0.2)(9.0,7.7)
      \rput[br](8.0,1.1){%
        \includegraphics[width=9.5cm]{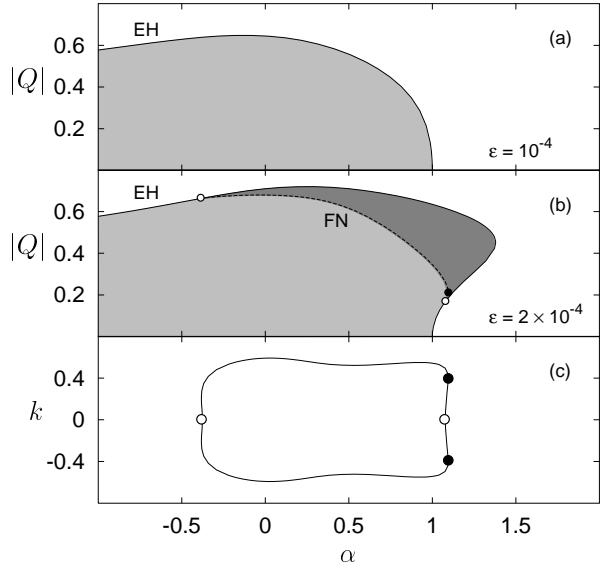}%
        }
    \end{pspicture}
    \caption{%
      (a) and (b) show curves in the ($\abs{Q},\alpha$) plane
      describing Eckhaus (EH) and finite-wavelength (FN)
      instabilities of plan wave solutions of the
      DSHE~\eno{eq:DSHE} for two different values of
      $\epsilon$ and $\beta = -1$.  (c) describes the
      variation along the FN curve in (b) of the marginally
      unstable wavenumber of the finite perturbation.
      }%
    \label{fig:StabDiagramTW}
  \end{center}
\end{figure}

To investigate the stability of the plane
waves~\eno{eq:PlaneWave}, we consider the growth rate
$\sigma(k)$ of longitudinal perturbations with $\vec{k}
\parallel \vec{Q}$.  For the DSHE, an analytic evaluation of
the spectrum of eigenvalues requires the solution of a cubic
polynomial with complex coefficients, and is therefore not
suitable for analytic evaluation.  Instead, we may apply
second order linear perturbation theory \cite{Kato} to obtain
a series expansion for the growth rates: For the first order
correction to the Eckhaus criterion~\eno{eq:Eckhaus}, we
obtain for the DSHE
\begin{equation}
  E_\parallel = D_\parallel +
  \frac{4(\alpha-\beta)(\alpha - \gamma)\kappa Q^2}%
  {\lambda_0(\lambda_0 + \kappa)} \epsilon = 0,
  \label{eq:EckhausDshe}
\end{equation}
where $D_\parallel$ is the Eckhaus criterion~\eno{eq:EigCGLE}
for the CGLE.  For $Q=0$, we observe that the BFN criterion
also holds for the DSHE.  For the CGLE, all plane waves are
long-wavelength unstable when $1+\alpha\beta < 0$.  For the
DSHE, this no longer holds, since a band of plane waves of
finite wavenumber still remain stable at the BFN point when
$\epsilon > -\lambda_0(\lambda_0 + \kappa)/(2\kappa(1 +
\beta\gamma))$.  This band of plane waves, however, can become
unstable to finite-wavelength perturbations determined by the
condition
\begin{equation}
  F_\parallel = \re\sigma'(k) = 0, \qquad \abs{k} > 0.
  \label{eq:ConvCrit}
\end{equation}
For example, for a homogeneous plane wave ($Q = 0$), expansion
of $\sigma(k)$ to lowest nontrivial order in $\epsilon$ and
fourth order in $k$ yields
\begin{align}
  \sigma(k) = 
  - (1+\alpha\beta)k^2 
  \pm \tfrac{1}{2}
  \bigl[(1+ \alpha^2)\beta^2 
  + 2\frac{\beta( 1+\alpha\beta-\delta)(\alpha -\gamma)\kappa}
  {\lambda_0(\lambda_0 + \kappa)}\epsilon\bigr]k^4.
  \label{eq:DsheEig}
\end{align}
For the CGLE, the long-wavelength instability will always be
the first plane wave instability to occur for the plane wave
with $Q=0$.  However for the DSHE this result no longer holds,
since the coefficient of order $k^4$ in \eqn{eq:DsheEig}
changes sign when
\begin{equation}
  \epsilon = 
  \frac{(1+ \alpha^2)\beta\lambda_0(\lambda_0 + \kappa)}
  {2(1+\alpha\beta-\delta)(\alpha-\gamma)\kappa}
  \label{eq:EpsCrit}
\end{equation}
causing a finite-wavelength instability to take place.  A
similar result can be derived for general $Q$.  A typical band
of unstable wave numbers are shown in \fig{fig:Eigval}.  We
also observe that \eqns{eq:EckhausDshe}{eq:DsheEig} converges
to the corresponding stability results of the CGLE in either
of the adiabatic limits $\epsilon \rightarrow 0$ or $\lambda_0
\rightarrow -\infty$.  The described scenario is illustrated
in \fig{fig:StabDiagramTW}.  In (a), $\epsilon = 10^{-4}$, we
obtain qualitatively the same behavior as in the CGLE, where
the band of Eckhaus stable plane waves vanish at the BFN
point.  In (b), $\epsilon = 2.0 \sci{-4}$, a finite-wavelength
band now persists above the BFN point.  In addition, a
finite-wavelength instability emerges at two points from the
Eckhaus curve (open circles on figure).  As $\alpha$
increases, the corresponding instability curve exhibits a
limit point (black circle); above this point, all plane waves
are unstable.  The variation along the finite instability
curve of the marginally unstable wavenumber of the finite
perturbation is shown in (c).

Similar to the CGLE, the DSHE admits spiral wave solutions
(phase defects) in both one and two spatial dimensions.  In
two spatial dimensions, these may be expressed in polar
coordinates $(r,\theta)$ as $w(r,\theta) = A(r)\exp(\I(\psi(r)
- \theta - \omega t))$ and $z(r) = Z(r)$, where $A(r)$ and
$\psi(r)$ are the amplitude and phase of the spiral wave
respectively in the complex $w$ component whereas $Z(r)$ is
the amplitude of the slow $z$ component.  These three
quantities must satisfy the boundary value problem
\begin{subequations}
  \begin{align}
    A(0) &= \psi(0) = Z(0) = 0, \\
    \lim_{r\rightarrow\infty} A(r) &= 
    \sqrt{(1-\qs^2)/(1 + \kappa/\lambda_0)}, \\
    \lim_{r\rightarrow\infty} Z(r) &= -A(r)^2 \kappa/\lambda_0
  \end{align}
  \label{eq:DsheBVP}%
\end{subequations}
where $\qs = \lim_{r\rightarrow\infty} \psi'(r)$ is the unique
wave number selected by the spiral wave.  We now discuss the
stability properties of spiral wave solutions of
\eqn{eq:DsheBVP}. In order to compare the results with the
properties of the CGLE, we shall use the parameter $\alpha$
determined implicitly by \eqn{eq:AlphaDSHE} as the free
parameter, whereas all other parameters in the
DSHE~\eno{eq:DSHE} are kept fixed.

For large $r$, spiral waves resemble plane wave solutions of
the form \eqn{eq:PlaneWave}, and we may therefore expect that
the spiral wave stability is governed by the corresponding
stability for a plane wave with $Q = \qs$.  The transitions
where spiral waves become either Eckhaus unstable or unstable
to finite-wavelength perturbations are therefore
$D_\parallel(\qs) = 0$ or $F_\parallel(\qs) = 0$ respectively.
However, as described in \cite{Ar92a}, spiral waves emit plane
waves with a nonzero group velocity $\im \partial
\sigma/\partial k$, causing perturbations to drift away; the
conditions~\eno{eq:EckhausDshe} and \eno{eq:ConvCrit} are
conditions for convective instability, and can therefore only
be taken as necessary criteria for instability.  To determine
exponential growth of a perturbation $u(\vec{x}, t)$ even in a
steady coordinate frame, we must evaluate the Fourier integral
\begin{equation}
  u(\vec{x},t) = \frac{1}{2\pi}\int_{-\infty}^{\infty}
  u_k(0)\E^{\I k x + \sigma(k)t} dk
\end{equation}
for large $t$ in the saddle-point
approximation~\cite{Mathews70}.  The crossing to absolute
instability is then determined by the two conditions
$\sigma'(k_0) = 0$ and $\re \sigma(k_0) = 0$.

\begin{figure}[htbp]
  \begin{center}
    \begin{pspicture}(0,0)(9.0,7.0)
      \rput[br](9.0,0.9){%
        \includegraphics[width=10.0cm]{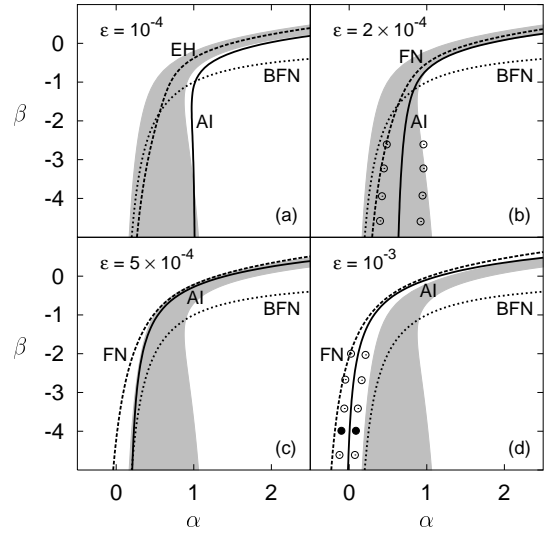}%
        }
    \end{pspicture}
    \caption{%
      Parameter diagrams showing the variation of the
      instability boundaries for spiral wave solutions of the
      DSHE~\eno{eq:DSHE} for four different values of the
      parameter $\epsilon$.  The figure shows the dominant
      boundaries for convective instabilities (dashed line),
      Eckhaus (EH) in (a) and finite-wavelength (FN) in (b-d),
      together with the boundary for absolute instabilities
      (AI, solid line).  The left and right boundaries of the
      gray-shaded area indicate the Eckhaus and absolute
      instability curves for the CGLE respectively.  The
      Benjamin-Feir-Newell line (BFN) is also shown.  For (b)
      and (d), small circles indicate points where the
      behavior has been confirmed by direct simulation of the
      DSHE.
      }%
    \label{fig:StabDiagram}
  \end{center}
\end{figure}

For four selected values of $\epsilon$, we have determined the
variation in the $(\alpha,\beta)$-plane of the DSHE
instability thresholds for Eckhaus, finite-wavelength, and
absolute instability as shown in \fig{fig:StabDiagram}.  To
solve the associated highly unstable boundary value problem,
we have used a continuation tool with support for multiple
shooting~\cite{Marek91}.  The corresponding Eckhaus and
absolute instability borders for the CGLE are also shown as
indicated by the gray-shaded area.  Even for $\epsilon$ small
(a), the Eckhaus threshold deviates significantly from the
corresponding CGLE curve whereas the absolute instability
curve almost coincides with the CGLE result.  Note that the
finite-wavelength instability does not exist for this value of
$\epsilon$.  However, as $\epsilon$ is increased (b-d), the
finite-wavelength curve completely determines the onset of
convective instability, and the Eckhaus curve has been omitted
from these panels.  We observe that both of the DSHE
instability limits gradually are shifted to the left of the
CGLE boundary.  Finally, for $\epsilon = 10^{-3}$ (d), the
instability limits lie completely outside of the limits
predicted by the CGLE.

\begin{figure}[htbp]
  \begin{center}
    \begin{pspicture}(0.0,0.0)(9.5,4.3)
      \rput[bl](-0.4,0.5){%
        \includegraphics[width=8.7cm]{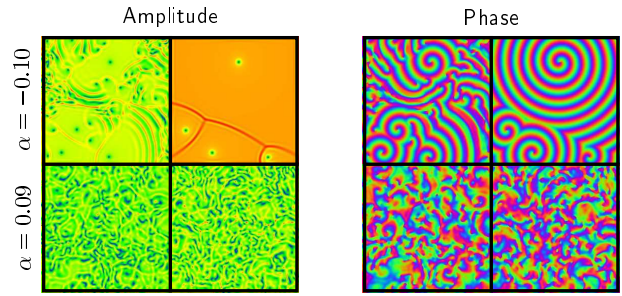}%
        }
    \end{pspicture}%
    \caption{%
      Snapshots showing the behavior of the DSHE near the
      onset of absolute instability corresponding to the two
      parameter points indicated by filled circles in the
      bifurcation diagram in \fig{fig:StabDiagram}.  For
      $\alpha = -0.10$, a convectively unstable transient ends
      in a frozen spiral state while $\alpha = 0.09$ gives
      rise to persistent spatio-temporal chaos.
      }%
    \label{fig:DsheSnap}
  \end{center}
\end{figure}

As observed for the CGLE\cite{Ar92a,CM96}, the absolute
instability (AI) line, is indicative for the onset of
persistent turbulence in the CGLE.  Numerical simulations of
the DSHE indicated by dots in
\figab{fig:StabDiagram}{fig:StabDiagram}{b}{d} confirms a
similar observation: spiral waves are convectively unstable
below the AI line and absolutely unstable above the AI line,
where a transition to sustained turbulence is observed.  A
representative scenario for the DSHE close to the AI line in
\figa{fig:StabDiagram}{d} is shown in \fig{fig:DsheSnap}.

The results derived in the stability analysis presented for
the DSHE~\eno{eq:DSHE}, show that the presence of a slow mode
in oscillatory chemical reaction-diffusion systems, can give
rise to a finite-wavelength instability of plane waves and
spiral waves, which does not occur in the CGLE.  In a real
chemical system, this instability occurs at a value of
$\epsilon$ where the amplitude of the oscillations is just
below the limit of detection.  So even close to the Hopf
bifurcation point, this instability completely determines the
stability of plane waves (which for the CGLE is given solely
by the Eckhaus criterion~\eqn{eq:Eckhaus}).  As shown in both
\figs{fig:StabDiagram}{fig:StabDiagramTW}, the
finite-wavelength instability has profound effects on the
location of boundaries for convective and absolute stability
for spiral waves, and completely alters the classical
bifurcation diagram known for the CGLE as $\epsilon$ is
increased.

For simple model systems of oscillatory chemical
reaction-diffusion systems, such as the Brusselator
\cite{BrusOrig} and the Gray-Scott model\cite{GS85}, the CGLE
provides an almost quantitative description of spatio-temporal
structures even at quite large distances from the bifurcation
point; however, models of realistic chemical and biochemical
systems, such as the BZ-reaction, the horseradish peroxidase
system~\cite{Aguda91}, and glycolytic
oscillations~\cite{Ni98,Ri75} all posses one or more slow
modes, and it is therefore unlikely that the CGLE will be
applicable for modeling experimental observations on such
systems.

We thank Igor Schreiber for valueable assistance with the
calculations of the stability boundaries shown
in~\fig{fig:StabDiagram}.

\bibliography{dynamic} 

\begin{thebibliography}{10}

\bibitem{Marsden76}
J.~E. Marsden and M. McCracken, {\em The Hopf Bifurcation and Its Applications}
  (Springer-Verlag, New York, 1976).

\bibitem{FieldBurger}
R.~J. Field and M. Burger, {\em Oscillation and Travelling Waves in Chemical
  Systems} (Wiley-Interscience, New York, 1985).

\bibitem{Kuramoto}
Y. Kuramoto, {\em Chemical Oscillations, Waves, and Turbulence}
  (Springer-Verlag, Berlin, 1984).

\bibitem{Ka95}
R. Kapral, {Physica~D} {\bf 86},  149  (1995).

\bibitem{Trans98}
M. Ipsen, F. Hynne, and P.~G. S{\o}rensen, to appear in \emph{Physica~D}
  (unpublished).

\bibitem{Wang95}
J. Wang, P.~G. S{\o}rensen, and F. Hynne, Z.\ Phys.\ Chem. {\bf 192},  63
  (1995).

\bibitem{New74}
A.~C. Newell, Lectures in Appl.\ Math {\bf 15},  157  (1974).

\bibitem{Ipsen97}
M. Ipsen, F. Hynne, and P.~G. S{\o}rensen, Int.\ J.\ Bifurcation and Chaos {\bf
  7},  1539  (1997).

\bibitem{GuckHolmes}
J. Guckenheimer and P.~J. Holmes, {\em Nonlinear Oscillations, Dynamical
  Systems and Bifurcations of Vector Fields} (Springer-Verlag, New York, 1983).

\bibitem{Kato}
T. Kato, {\em Perturbation Theory for Linear Operators} (Springer-Verlag,
  Berlin, 1966).

\bibitem{Ar92a}
I.~S. Aranson, L. Aranson, L. Kramer, and A. Weber, Phys.\ Rev.\ A. {\bf 46},
  R2992  (1992).

\bibitem{Mathews70}
J. Mathews and R.~L. Walker, {\em Mathematical Methods of Physics} (W. A.
  Benjamin, Inc., New York, 1970).

\bibitem{Marek91}
M. Marek and I. Schreiber, {\em Chaotic Behavior of Deterministic Dissipative
  Systems} (Cambridge University Press, Cambridge, 1995).

\bibitem{CM96}
H. Chate and P. Manneville, Physica~A {\bf 224},  348  (1996).

\bibitem{BrusOrig}
R. Lefever and I. Prigogine, J.\ Chem.\ Phys. {\bf 48},  263  (1968).

\bibitem{GS85}
P. Gray and S.~K. Scott, J.\ Phys.\ Chem. {\bf 89},  22  (1985).

\bibitem{Aguda91}
B.~D. Aguda and R. Larter, J.\ Am.\ Chem.\ Soc. {\bf 113},  7913  (1991).

\bibitem{Ni98}
K. Nielsen, P.~G. S{\o}rensen, F. Hynne, and H.-G. Busse, Biophysical Chemistry
  {\bf 72},  49  (1998).

\bibitem{Ri75}
O. Richter, A. Betz, and C. Giersch, BioSystems {\bf 7},  137  (1975).

\end{thebibliography}
\bibliographystyle{prsty}

\end{document}